\documentstyle[prl,aps,floats,epsf,epsfig]{revtex}

\begin{document}

\draft

\title{Stripe fractionalization II: the quantum spin nematic and the
Abrikosov lattice}

\author{J. Zaanen and Z. Nussinov}
\address{Instituut-Lorentz for Theoretical Physics, Leiden University\\
P.O.B. 9506, 2300 RA Leiden, The Netherlands}
\date{\today  
; E-mail: jan@lorentz.leidenuniv.nl}

\twocolumn[

\widetext
\begin{@twocolumnfalse}

\maketitle

\begin{abstract}

In part (I) of this two paper series on stripe fractionalization, we argued 
that in principle the `domain wall-ness' of the stripe phase could
persist in the spin and charge disordered superconductors, and we
demonstrated how this physics is in one-to-one correspondence 
with Ising gauge theory. Here we focus on yet another type of order
suggested by the gauge theory: the quantum spin nematic. Although
it is not easy to measure this order directly, we argue that the
superconducting vortices act as perturbations destroying the gauge
symmetry locally. This turns out to give rise to a 
simple example of a gauge-theoretical phenomenon known as
topological interaction. As a consequence, at any finite vortex density
a globally ordered antiferromagnet emerges. This offers a
potential  explanation for recent observations in the underdoped 214
system.

\end{abstract}

\vspace{0.5cm}

\narrowtext

\end{@twocolumnfalse}
]

\section{Introduction}

Among others, stripe order means that the charge stripes are domain
walls in the stripe antiferromagnet. In part I of this series of
two papers\cite{strifractI} we explained 
that the physics of this domain wall-ness in the case that the stripes form
a quantum liquid is formalized in terms of the most elementary 
field theory controlled by local symmetry: the Ising gauge theory.
We showed that the gauge fields have a geometrical meaning. These
parametrize the fluctuations of sublattice parity, the property that
a bipartite space can be subdivided in two ways in two sublattices:
$\cdots - A - B -\cdots$ or $\cdots - B - A -\cdots$. In  stripe
language, the ordered (deconfining) state of the gauge theory corresponds
with the stripes being intact as domain walls, implying that space is
either  $\cdots - A - B -\cdots$ or $\cdots - B - A -\cdots$. 
The theory predicts a phase transition corresponding with the destruction
of the stripe domain wall-ness, such that space turns non-bipartite 
(confinement). Remarkably, the gauge theory insists that this is a 
garden-variety quantum phase transition, which could be behind the
quantum criticality of the optimally doped cuprate superconductors.

We concluded part I with the observation that this topological (dis)order
can only be probed directly by topological means: non-local, multipoint
correlation functions (Wilson loops) which seem to be out of reach of
even the most fanciful experimental machine. At the same time, direct
experimental evidence is required because theoretically one can only
argue that it can happen. If it happens is a matter of microscopic
details, which cannot be analyzed in general terms. This part II is
dedicated to a potential way out of this problem. According to the
theory there is yet another state of matter to be expected: the
quantum spin nematic. This  corresponds with a superconductor carrying
a special type of anti-ferromagnetism characterized by an staggered 
order parameter which is minus itself (section II). Although such an
order cannot be observed by the standard probes of anti-ferromagnetism
(like neutron scattering and magnetic resonance) it is
not as hidden as the pure topological order of part I. 

By principle, superconducting order is required to protect the local
Ising symmetry. In the type II state of the superconductor, the
superconducting order is destroyed locally, in the vicinity of the
vortices. Accordingly, the vortices correspond with `gauge defects'
where the local Ising symmetry turns into a global one in isolated
regions in space. These gauge defects are quite interesting theoretically:
they correspond with an elementary example of the principle of `topological
interaction', non-dynamical influences mediating information over
infinite distances (section III). In the stripe interpretation this just
means that at the moment that vortices appear a piece of the spin-nematic
turns into a long range ordered anti-ferromagnet. In the final section
we give a recipe to study experimentally the spin nematic, 
making the case that it might well be
that the recently observed magnetic field induced antiferromagnet
in the $La_{1.9}Sr_{0.1}CuO_4$\cite{lake} is of this kind.

\section{The quantum spin nematic}

In part I, we assumed implicitly that both the antiferromagnetic order
and the charge order of the stripes were both fully destroyed and we
discussed the fluctuating domain wall-ness in isolation. However, there
is yet another state possible\cite{philmag,sachdev1,zhou}. 
As long as the stripe dislocations do
not proliferate, the spin system is not frustrated in essential ways;
it can be argued that the domain wall-ness of the static stripes has 
everything to do with organizing the motions of the holes in such a 
way that the frustrating effect of the isolated hole motions are 
avoided. This unfrustrating influence of the stripes stays intact
even when the stripes are completely delocalized, as long as they form
connected domain walls. Hence, a state can exist in principle where
the charge is disordered while next to the sublattice parity also the
spin system maintains its antiferromagnetic order. However, due to
the stripe fluctuations this is not a normal antiferromagnetic but
instead a spin-nematic.

\begin{figure}
\includegraphics[width=6.9cm]{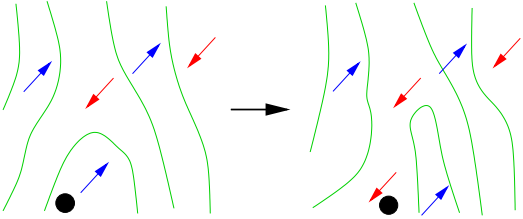}
\caption{ The physical nature of the quantum spin-nematic (or 
Higgs phase of the $O(3)/Z_2$ theory) in terms of fluctuating stripes.
As long as the domain walls are fully connected, the spin system 
is protected against frustration and one would see an antiferromagnetic
order (arrows) upon taking a snapshot on a time scale short compared
to the charge fluctuation scale. However, at long times the domain
walls are delocalized with the effect that the fluctuating stripes 
turn the staggered order parameter into minus itself: watch what 
happens at the black dot.}
\label{fig1}
\end{figure}

The nature of this state is easy to understand. Take a snapshot on
a timescale short as compared to the charge fluctuations and we
would see an ordered antiferromagnet except for the fact that the
staggered order parameter flips every time a domain wall is crossed
(fig. 1). At some later time it
will look similar except that all domain walls will have moved. At
long times, we cannot say where the domain walls are with the ramification
that the staggered order parameter becomes minus itself:
$\langle \mbox{\boldmath$M$}  (\mbox{\boldmath$r$}) \rangle 
\equiv \langle (-1)^{\mbox{\boldmath$r$}} 
\mbox{\boldmath$S$} (\mbox{\boldmath$r$}) \rangle \equiv - \langle 
\mbox{\boldmath$M$} (\mbox{\boldmath$r$})\rangle$.
Hence, the order parameter is no longer a $O(3)$ vector but instead 
an object pointing on the sphere having no head or tail: this is the
director (or `projective plane') order parameter well known from nematic
liquid crystals, and it is therefore called a spin nematic\cite{andreev}.

This can be easily formalized in terms of a gauge theory\cite{toner}.
The (fluctuating) antiferromagnetic order can be described in terms of
(coarse) grained $O(3)$ quantum rotors $\mbox{\boldmath$n$}$, 
quantized by an angular
momentum $\mbox{\boldmath$L$}$, 
such that $[ L^{\alpha}, n^{\beta} ] = i 
\varepsilon^{\alpha \beta \gamma} n^{\gamma}$. As compared to the
usual quantum non-linear sigma model description, the only difference 
is that the rotors are now minimally coupled to the $Z_2$ gauge fields.
We remind the reader of the Hamiltonian of the pure Ising gauge
theory\cite{Kogut}, 
parametrizing the dynamics of the domain wall-ness (see part I),
\begin{equation}
H_{gauge}  =  -K \sum_{\Box} \sigma^3 \sigma^3 \sigma^3 \sigma^3
- \sum_{<ij>} \sigma^1_{ij}
\label{Z2gauge}
\end{equation}
where $\sigma^{1,3}$ are Pauli-matrices acting on Ising bond variables.
$\sum_{\Box} \sigma^3 \sigma^3 \sigma^3 \sigma^3$ is the plaquette
interaction, such that Eq. (\ref{Z2gauge}) commutes with the
generator of gauge transformations $P_i = \Pi_j \sigma^1_{ij}$. To couple
in the matter fields, put the rotors on the sites of the lattice of
the gauge theory, and define
\begin{equation}
H_{O(3)/Z_2} = H_{gauge} - J \sum_{<ij>} \sigma^3_{ij} \mbox{\boldmath$n$}_i
\cdot \mbox{\boldmath$n$}_j - \sum_i L^2_i
\label{O(3)/Z2}
\end{equation}
Hence, the gauge fields determine the sign (`ferro' or `antiferromagnetic')
of the `exchange' interactions between the rotors on neighboring sites. 
Consider the case
that both $K$ and $J$ are large. The gauge sector will be deconfining and the
unitary gauge fix (all bonds +1) is representative\cite{Kogut}. 
Since $J$ is also
large the $O(3)$ symmetry is also spontaneously broken
 and  all rotors will point in 
the same direction (Fig. 2). Apply now a gauge transformation at some site $i$;
all bonds emerging from this site will turn from ferromagnetic in
antiferromagnet and when one multiplies simultaneously 
$\mbox{\boldmath$n$}_i$ by $-1$ 
the energy will stay invariant. Hence, the gauge transformations take care
of changing the (unphysical, non gauge invariant) antiferromagnet into the
physical (gauge invariant) spin-nematic, characterized by a staggered
order parameter `having no head or tail' (actually, the projective plane).
Eq. (\ref{O(3)/Z2}) is just
the quantum interpretation of the classical $O(3)/Z_2$ model studied
in a great detail Lammert, Rokhsar and Toner\cite{toner}. 
The phase diagram
is completely known,  and the spin disordered deconfining and 
confining phase discussed in part I share a second
order 3D Heisenberg transitions- and a first order quantum phase transition
with the spin nematic, respectively.

\begin{figure}
\includegraphics[width=7.3cm]{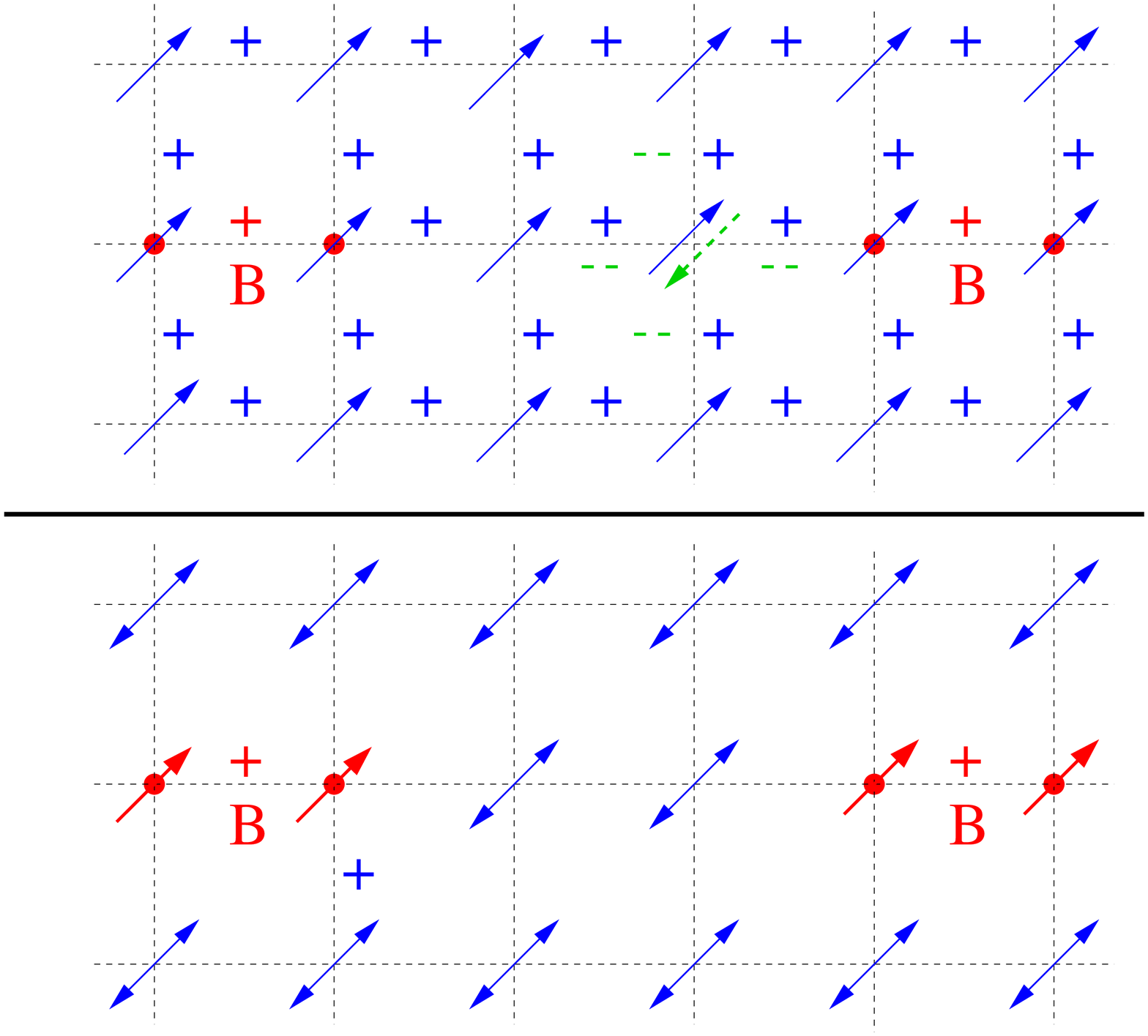}
\caption{Construction of the spin nematic state, and the topological 
interactions between the spatially disconnected gauge defects. The
unitary gauge (all bonds +) is representative and for large $J$ the
rotor degrees of freedom living on the sites will also order (upper
panel). By performing gauge transformations (the dashed  bonds and spins)
the rotors turn into directors, which are like vectors except that 
their heads and tails are the same (lower panel). By applying an
external field $B$ giving a definite sense to the sign of isolated
bonds the gauge symmetry is broken at the 4 sites labeled by
dots in the figure. Remarkably, one finds following the same procedure as
for in the absence of the gauge symmetry breaking that the heads of 
the $O(3)$ vectors at the gauge defects all point in the same direction.}
\label{fig2}
\end{figure}

Could there be such a spin nematic phase around in the context of
cuprate superconductors? An obvious place to look for it would
be the underdoped 214 system with its strong tendency towards 
antiferromagnetic order. In highly doped 2212 and 123 there are
good reasons to believe that for other reasons the spin system
is strongly quantum disordered. The spin nematic shares the attitude
with the domain wall gauge fields to hide itself from detection 
in standard experiments. However, it is not as successful in this
hiding game as the pure gauge fields are. Antiferromagnets can
be directly probed using neutron scattering, NMR and $\mu$SR, because
these experiments measure in one or the other way the two point
(staggered) spin correlator $S(| \mbox{\boldmath$r$} - 
\mbox{\boldmath$r$}'|) = \langle \mbox{\boldmath$M$} (\mbox{\boldmath$r$})
\mbox{\boldmath$M$} (\mbox{\boldmath$r$}') \rangle$. Because in the
spin nematic
$\mbox{\boldmath$M$} (\mbox{\boldmath$r$})
\equiv - \mbox{\boldmath$M$} (\mbox{\boldmath$r$})$,  independently
at every $\mbox{\boldmath$r$}$, it follows that  $S = - S$,
 meaning that it has to vanish: $S$
is not gauge invariant.  Employing again the `stripe detectors' of part I
($\tilde{\sigma}^3 (\mbox{\boldmath$r$})$ acquiring values $-1$, $+1$ when
a domain wall is detected or not, respectively), 
the gauge invariant correlation function which
can `see' the spin nematic order is  $S_{Z2} (| \mbox{\boldmath$r$} - 
\mbox{\boldmath$r$}' |)
= \langle \mbox{\boldmath$M$} (\mbox{\boldmath$r$})
\Pi_{\mbox{\boldmath$l$}=\mbox{\boldmath$r$}}^{\mbox{\boldmath$r$}'}
\tilde {\sigma}^3 (\mbox{\boldmath$l$}) ) \mbox{\boldmath$M$} 
(\mbox{\boldmath$r$}')\rangle$, i.e. the
`matter correlator with the Wilson line inserted'. Relative to
the Wilson loops of part I, this does not seem to add much to the
comfort of the experimental physicist.

However, with the matter fields present there is more to look for.
 In the coarse grained $O(3)$ 
language, although $\mbox{\boldmath$n$}$ is not gauge invariant the traceless
tensor $Q_{\alpha \beta} = n^{\alpha} n^{\beta} - 1/3 \delta_{\alpha
\beta}$\cite{andreev,toner} 
is a gauge singlet because it transforms like $n^2$. This
tensor is actually measured in two magnon Raman scattering\cite{kivelson}.
There is unfortunately a practical problem. Imagine that a spin nematic
would be realized in, say, $La_{2-x}Sr_x CuO_4$. The 5 meV gap observed
in the superconducting state in the spectrum of incommensurate spin
fluctuations would then be interpreted as the charge fluctuation scale.
At energies below the gap the structure factor vanishes because the spin 
nematic sets in. However, at energies above the gap  
the antiferromagnetism becomes visible 
because the neutrons are just  `taking the snapshots' as in Fig. 1.
On a side, this interpretation actually offers
a simple interpretation for the observation that this gap disappears 
above the superconducting $T_c$: when the phase order disappears the
charge fluctuations become relaxational and there is no longer a 
characteristic charge fluctuation scale protecting the gauge invariance
dynamically\cite{ZaHovS,Zaphysica},
 although it might be still around in the statics\cite{koeln}.
In order to nail down the spin nematic one would like to see the
characteristic behavior associated with spin waves in the Raman response
(intensity $\sim \omega^3$) at energies less than 5 meV where the neutrons 
seem to indicate there is nothing. Unfortunately it seems impossible
to isolate the two magnon scattering from the Raman
signal at these low energies\cite{blumberg}.

\section{Vortices as gauge defects}

Fortunately, there is a much less subtle way to look for the spin nematic.
As we explained in part I, the emergence of the gauge invariance requires 
the presence of the superconducting order. Hence, when superconductivity
is destroyed the gauge invariance is destroyed and the local $Z_2$ symmetry
turns global. Upon applying a magnetic field to the superconductor, the
Abrikosov vortex lattice is created where the superconductivity is locally
destroyed in the vicinity of the vortices. This suggests that we have
to consider the general problem of what happens with the gauge theory 
when the gauge invariance turns into global $Z_2$ invariance at isolated
regions in space: the `gauge defects'. Let us first consider this problem 
on an abstract level, using the lattice gauge theory, to continue 
thereafter with a consideration what this all means for stripes.

Breaking the gauge symmetry, even
in isolated spots in space, is a brutal operation. In first instance
 it does not matter how one breaks it. Let us therefore take the Hamiltonian  
Eq.'s(\ref{Z2gauge},\ref{O(3)/Z2}) and add the simplest `impurity' term 
breaking  the local symmetry,
\begin{equation}
H_{imp} = -  B \sum_{<kl>}  \sigma^3_{kl}  
\label{gaugeimp}
\end{equation}
where we pick some bonds $kl$ as the `impurity sites'. This term `removes'
the gauge from the bond, and the gauge invariance is destroyed on
the two sites connected by the $kl$ bond when $B \neq 0$. 
For a single impurity,
the symmetry turns locally into a global $O(3)$ symmetry. Consider now
the case that spin nematic order is present and insert two gauge defects
with $B > 0$, separated by some large distance (Fig. 2). Take the
unitary gauge: all bonds $+1$ including the impurity bonds. Obviously,
when $K$ and $J$ are both large this is a representative gauge, regardless
the presence of the two $+1$ global bonds, and in this gauge all rotors 
point in the same direction. In a next step, perform gauge transformations
everywhere
except for the four sites where the gauge symmetry broken. This will turn
the medium into a spin nematic (Fig. 2). What has happened? Although the two
impurity sites are separated by a medium which seem to have no knowledge
about where the heads and the tails of the rotors at the impurity site
are, there seems to be a remarkable `action at a distance': although
the two impurity sites can be infinitely apart the spins know that they
have to stick their heads in the same direction! It is easily checked 
that the unitary gauge stays representative also in the presence of
virtual vison pairs and it is only when the visons proliferate, 
destroying the spin nematic order, that this `action at a distance' 
is destroyed. The conclusion is that a local breaking of the gauge
invariance suffices to cause an  global $Z_2$ `headness' 
{\em long range order} of the rotors, so that they together break
the ungauged $O(3)$ symmetry.
In this sense the local symmetry is infinitely
fragile with regard to global violations. It is noticed that the
above is an elementary example of a topological interaction, i.e.
an information carrying influence which is entirely non-dynamical
and not mediated by propagating excitations.  These
are known to occur in much less trivial theories, like for
instance 2+1 dimensional gravity\cite{qugrav}. 

In fact, the above is not quite representative yet for the stripe case,
because we have to build in communication with the translational 
symmetry. All we have in the gauge theory is the simple `auxiliary'
lattice on which the theory is defined, and the minimal way to
let the spin system know about this lattice is by incorporating
a sense of antiferromagnetism. Upon breaking the gauge this is easily
achieved by taking for the gauge defects  a negative `exchange' $B < 0$.
The `action at a distance' for this case can be constructed in a similar
way as for the `ferromagnetic' case. 
Start again with unitary gauge (everywhere +1 bonds) and perform 
gauge transformations producing a negative bond at the impurity bonds,
to subsequently restore the gauge invariance away from the impurity
sites. One now encounters an ambiguity. One can perform the gauge 
transformation on the site to the `left' or the `right' of a impurity
bond, and one finds that pending this choice the orientation
of the staggered order reverses relative to a reference impurity. At
first sight it seems that for staggered configurations the `action at a
distance' fails, because the heads and the tails of the local 
staggered order parameters point in arbitrary directions. 
However, this is not the case: this indeterminedness
has nothing to do with the `topological gauge force' but instead with
a left-over translational invariance. The generators of gauge transformations
live on the sites and by breaking the gauge invariance on a single bond,
the gauge invariance is broken on the two sites connected by this bond
which remain therefore translationally equivalent. This translation
is responsible for the flipping of the staggered order. One should 
instead center the gauge symmetry breaking on a site. Apply for instance
the symmetry breaking $- B \Pi_l \sigma^3_{kl}$, fixing all 
bonds coming out of the site $k$, to find that in this case the gauge
action-at-a-distance acts in exactly the same way for the staggered
order parameter as it does for the uniform case.

Summarizing, using an elementary argument, we identified a ghostly,
non-dynamical action at a distance ordering the rotors at spatially 
disconnected  `gauge impurities' which requires nothing more than
spin nematic order. As a caveat, we found that in order to find the
same global order for staggered spin we have to add as an extra
requirement also the translational symmetry breaking by the impurities.
We will now argue that these general features of the gauge theory
acquire a quite mundane interpretation in terms of the stripes.

\section{Magnetic field induced antiferromagnetism}

Anything in the gauge theory should be in one-to-one correspondence
to something in stripe physics. This is also true for the gauge defects
and in fact it becomes so simple in the stripe interpretation that the latter
is an ideal tool to convince the gauge theory student that the ghostly
`action at a distance' is actually not a big deal.

Given that the spin nematic exist, it has to be that the competitor
of the superconductor is a fully ordered stripe phase. As we will
discuss in more detail, it is reasonable to expect
that in the proximity of the vortex cores 
the charge density order of the stripe phase will
re-emerge, and it might be that this is already observed in the form of
the stripy `halo's' surrounding the vortex cores as seen by Hoffman and
coworkers by STM\cite{davis}. 
Charge is bound to the domain wall-ness and when charge
orders the domain walls come to rest, and
the spin-nematic turns into a stripe antiferromagnet which can be seen
by conventional means like neutron scattering, see Fig. 3. 
The charge order is the
gauge defect, making the magnetic order visible which already 
pre-existed in the superconductor. The amount of antiferromagnetic order
is expected to be proportional to the volume taken by the charge-ordered
halo's, because this corresponds to the volume of the system where the
local symmetry turned global. What determines the correlation length
of this antiferromagnet? We remind the reader of the translational
symmetry breaking required for staggered order as discussed in the
previous section. In the stripe context it has the following meaning.
Although the spins always become static, a full stripe antiferromagnetic
order also needs a full translational order of the sublattice parity 
which is the same as  translational order in
the charge sector: the antiferromagnetic correlation length 
is identical to the charge-order correlation length.

\begin{figure}
\includegraphics[width=7.1cm]{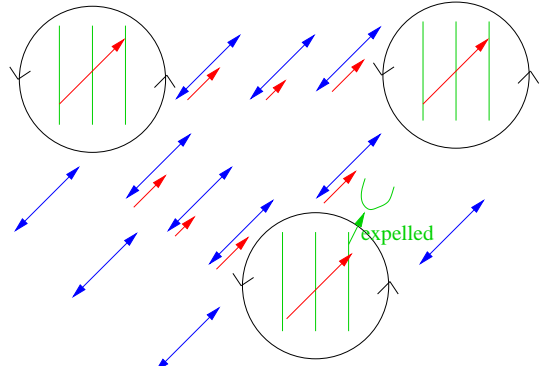}
\caption{ A cartoon picture of the charge coming to rest at the
vortex cores, with the effect that the sublattice parity stops
fluctuating locally. This automatically turns the spin nematic
into a normal anti-ferromagnet.}
\label{fig3}
\end{figure}

With regard to the thermal phase transition
one expects that the spin nematic behaves similar to the antiferromagnet.
In strictly 2D one cannot have true spin-nematic long range order 
(LRO)\cite{koeln}.
However, due to weak 3D couplings, etcetera, one expects nevertheless a
true LRO at low temperatures. A difficult question is if the spin nematic
completely disorders at this finite temperature transition
or that a topologically ordered phase can
be realized. In the first case, the transition has to be first order but
it is likely so weakly first order that it is hard to distinguish from
a second order transition. Our most striking  prediction is that when
an external magnetic field is applied, the temperature where this
 thermal phase transition occurs should at least initially be field
independent.  The reason is simple. In the absence of the field
the spin-nematic order is already well developed, protected by
a large cohesive energy of order of the observed transition
temperature $\sim 40$K. Since the external field couples in through
its energy, and since the field energy (a few Tesla's) is small compared
to the spin-nematic cohesive energy, the field cannot change the transition
significantly.   Hence, the 
specialty of the quantum spin nematic, which we believe is unique to this
form of matter, is that it causes
an apparent dissimilarity between the sensitivity of
the zero-temperature antiferromagnetic order as induced 
by the magnetic field and the insensitivity of the thermal phase transition 
temperature to the same field.
The magnetic order is already strongly developed at zero field but it 
cannot be measured by neutrons, etcetera. Upon applying the field, the
spin nematic turns in part into an antiferromagnet, 
becoming visible in magnetic experiments with a magnitude 
determined by the induced charge order.
This is to be strongly contrasted with the `conventional' interpretation
that the the magnetic field {\em creates} the antiferromagnetic order.

Zhang, Demler and Sachdev\cite{sachdev1}
have developed a general phenomenological
theory, dealing with the case that the antiferromagnetic order is
created by the field, arriving at a number of strong predictions.
Their starting point is a soft-spin, Ginzburg-Landau-Wilson description
of the antiferromagnetic order parameter field $\phi$ and the superconducting
field $\Psi$. The lowest order coupling between the two fields is 
$B |\Psi|^2 |\phi|^2$.  They arrive at the counter-intuitive
result that, starting with a quantum disordered antiferromagnet, one has
to exceed a critical strength of the magnetic field before LRO 
antiferromagnetism sets in which is delocalized over the system. 
The reason is the self-interaction of the antiferromagnetic order
parameter field preventing it from localizing itself in the vicinity
of the vortex cores. Comparing it to the data by Lake et al.\cite{lake}, 
they argue that $La_{2-x}Sr_x CuO_4$ shows already antiferromagnetic 
order in zero-field implying that this superconductor coexists with
an antiferromagnet. A worry is that this zero field antiferromagnetism 
has a completely different temperature dependence (not showing signs
of a  finite temperature phase transition) while it is apparently varying
strongly from sample to sample, suggesting that it is a dirt effect.
At the same time, the field induced antiferromagnetism seems to come
up smoothly with the field 
and there is no sign of a critical threshold. Even more
worrisome is the fact that the temperature where the 
 field induced antiferromagnetism appears is rather 
independent of the applied field and this is very hard to understand
in this competing order framework. Since the antiferromagnetic order
is created by the field, it is very weak when the field is small
and accordingly one would expect that initially $T_N$ is very small,
increasing rapidly with the increase of the zero temperature staggered
order parameter. In fact, assuming that $T_N$ is due to 3 dimensional
couplings and spin anisotropies, one expects $T_N$ to be linearly
proportional to $M_0$ \cite{duin}, the zero temperature staggered magnetization
for small $M_0$. Instead, $T_N$ is in the Lake experiments rather
field independent and we take this as strong evidence in favor of
the spin nematic (Fig. 4).

\begin{figure}
\includegraphics[width=7.15cm]{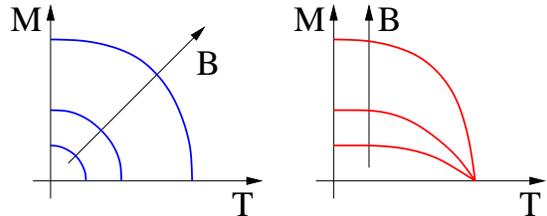}
\caption{Left panel: the expected magnetic field dependence of the
induced magnetism as function of temperature in the competing order
framework. The magnetic transition temperature is expected to be
strongly dependent on the applied field. Right panel: assuming 
that a spin-nematic is present, the transition temperature should
barely depend on the field, because the order is already present in
the absence of the field, to become just observable in the neutron
scattering in the presence of the field. The data of Lake et al.
for $La_{1.9}Sr_{0.1}CuO_4$ look like the right panel.}
\label{fig4}
\end{figure}

Can the fraction of the spin nematic turning into antiferromagnetic
order as function of the magnetic field be quantified? In fact, this
is possible although the solution is only available right now in
numerical form. The problem of the pinning of the charge density
wave by the vortex lattice is also addressed in some detail by
Zhang et al.\cite{sachdev1}. 
The crucial difference with the antiferromagnet is 
that the charge density wave communicates directly with the
vortex lattice because both fields break translational invariance.
As a result, the vortex-lattice acts as a spatially varying potential
on the charge-ordering field (Eq. (1.12) in ref.\cite{sachdev1}) with the
consequence that charge order directly accumulates in
the vicinity of the vortex cores at any value of the external field.
Zhang {\em et al.} present some numerical results on the behavior of the charge
order in the magnetic field (Fig. 15, 16 in ref.\cite{sachdev1}). 
A caveat is that these
are calculated in the presence of a low lying magnetic exciton and
it is not immediately clear if these results are directly applicable
to the spin nematic case. A related issue is to what extent
the commensuration effects associated with the stripe charge order 
versus the vortex lattice can give rise to strong charge order
correlations between the `halo's' centered at different vortices.
As we discussed, such correlations are a necessary condition to
find  correlations in the spin system exceeding the vortex distance.
Notice, however, that these theoretical difficulties can be circumvented
using  experimental information: when the spin nematic is
present, the antiferromagnet order should closely follow the
charge order, in strong contrast with the expectations following 
from the competing order ideas.

In conclusion, we have presented the hypothesis that in underdoped
$La_{2-x}Sr_xCuO_4$ a new state of quantum matter might be present:
a superconductor which is at the same time showing spin nematic order.
We have argued that it should be possible to proof or disproof the
presence of such a state using conventional experimental means, while
existing experiments already strongly argue in favor of this possibility.
What really matters is that, if the spin nematic is indeed realized, the
proof of principle is delivered that the domain wall-ness of the ordered
stripe phase can persist in the quantum fluid. This would add credibility
to the possibility that the stripe topological order could even persist
in the absence of any spin order, which in turn could be responsible for
the anomalies of the best superconductors.

Acknowledgments. We acknowledge helpful discussions with S. Sachdev, 
S.A. Kivelson, G. Aeppli, G. Blumberg, H. Tagaki,
B. Lake, F. Zhou, E. Demler and
especially P. van Baal for his comments on the topological interactions
in gauge theories. This work was supported by the Dutch Science 
Foundation NWO/FOM.



\begin{thebibliography}{9}


\bibitem{strifractI}
Z. Nussinov and J. Zaanen, proceedings of the international workshop
ECRYS 2002, to appear in J. Phys. (Paris) Coll. (cond-mat/xxxxxx)
\bibitem{lake}
B. Lake {\em et al.}, Nature {\bf 415}, 299 (2002).
\bibitem{philmag}
J. Zaanen, O.Y. Osman, H.V. Kruis, Z. Nussinov, and J. Tworzydlo,
Phil. Mag. B {\bf 81}, 1485 (2002).
\bibitem{sachdev1}
Y. Zhang, E. Demler, and S. Sachdev, Phys. Rev. B {\bf 66}, 094501
(2002) 
\bibitem{zhou}
Similar ideas have been explored in the context of Bose-Einstein condensates:
E. Demler and F. Zhou, Phys. Rev. Lett. {\bf 88}, 163001 (2002).
\bibitem{andreev}
The principle of spin-nematic order was introduced by
A.F. Andreev and I.A. Grishchuck, Sov. Phys. JETP {\bf 60}, 267 (1984).
\bibitem{toner}
P.E. Lammert, D.S. Rokhsar and J. Toner, Phys. Rev. Lett. {\bf 70}, 
1650 (1993).
\bibitem{Kogut}
J. B. Kogut, Rev. Mod. Phys. {\bf 51}, 659 (1979).
\bibitem{kivelson}
S.A. Kivelson, private communications.
\bibitem{ZaHovS}
J. Zaanen, M.L. Horbach and W. van Saarloos, Phys. Rev. B {\bf 53}, 10667 
(1996).
\bibitem{Zaphysica}
J. Zaanen and W. van Saarloos, Physica C {\bf 282}, 178 (1997).
\bibitem{koeln}
F. Kruger and S. Scheidl, Phys. Rev. Lett. {\bf 89}, 095701 (2002).
\bibitem{blumberg}
G. Blumberg, private communications.
\bibitem{qugrav}
S. Carlip, `Quantum gravity in 2+1 Dimensions', (Cambridge Univ. Press, 1998). 
\bibitem{davis}
J.E. Hoffman {\em et al.}, Science {\bf 295}, 466 (2002).
\bibitem{duin} C. N. A.  can Duin and J. Zaanen, Phys. Rev. Lett. {\bf
80}, 1513 (1998)


\end{thebibliography}
\end{document}